# Research on Deep Learning Model of Feature Extraction Based on Convolutional Neural Network

Houze Liu[1], Iris Li[2], Yaxin Liang[3], Dan Sun[4], Yining Yang[5], Haowei Yang[6]

[1]New York University, Department Computer Science and Engineering, New York, USA, hl2979@nyu.edu
[2]New York University, Graduate School of Art & Science, New York, USA, irisepode@gmail.com
[3]University of Southern California, Viterbi School of Engineering , Los Angeles ,USA, yaseen.liang@outlook.com
[4]Washington University, Department of Electrical and Systems Engineering, St. Louis, USA, sun.dan@wustl.edu
[5] Carnegie Mellon University, School of Computer Science , Pittsburgh, USA, yiningya@andrew.cmu.edu
[6]University of Houston, Cullen College Of Engineering, Houston, USA, yanghaowei09@gmail.com

*Abstract*—Neural networks with relatively shallow layers and simple structures may have limited ability in accurately identifying pneumonia. In addition, deep neural networks also have a large demand for computing resources, which may cause convolutional neural networks to be unable to be implemented on terminals. Therefore, this paper will carry out the optimal classification of convolutional neural networks. Firstly, according to the characteristics of pneumonia images, AlexNet and InceptionV3 were selected to obtain better image recognition results. Combining the features of medical images, the forward neural network with deeper and more complex structure is learned. Finally, knowledge extraction technology is used to extract the obtained data into the AlexNet model to achieve the purpose of improving computing efficiency and reducing computing costs. The results showed that the prediction accuracy, specificity, and sensitivity of the trained AlexNet model increased by 4.25 percentage points, 7.85 percentage points, and 2.32 percentage points respectively. The graphics processing usage has decreased by 51% compared to the InceptionV3 mode.

## I. Introduction

In the past diagnosis of pneumonia, the determination of images mostly relied on doctors with rich clinical practice experience, and this difference could not be guaranteed to be correct due to artificial experience. In recent years, computer-assisted imaging (CCAD) has been gradually applied in the imaging field [1], and has become an important basis for doctors to evaluate image quality [2]. Researchers have classified the images of pneumonia images and given the corresponding algorithms [3]. Researchers have used a 121-level convolutional neural network to perform experiments on 112,120 labeled lung X-ray images, and found that 11 of them achieved results comparable to or better than imaging diagnoses by radiographers [4].

In this paper, the convolutional neural network based on knowledge extraction is optimized to enhance the identification performance of pneumonia. AlexNet with the initial version was selected [5]. This algorithm uses V3 and AlexNet as research modules to study the transfer learning-based neural network model to solve the overfitting problem caused by the small number of samples [6]. Further improve AlexNet's recognition performance while reducing GPU computing resource consumption [7].

## II. Pneumonia identification and network compression methods

### A. Pneumonia classification network structure design

Firstly, this paper analyzes the current mainstream deep neural network architecture, including AlexNet, VGG16, ResNet18, ResNet34, InceptionV3, and so on [8]. The error rate, parameter number, network depth, and so on are compared [9].

TABLE I.　Comparison of performance of each model

| Model name | Top-1 error rate /% | Network depth | Number of parameters /$10^6$ | Computation /$10^6$ |
|---|---|---|---|---|
| AlexNet | 38.23 | 8 | 64.90 | 750 |
| VGG16 | 25.73 | 24 | 144.06 | 15938 |
| ResNet18 | 29.04 | 19 | 14.79 | 1875 |
| ResNet34 | 22.43 | 35 | 25.21 | 3750 |
| Inception V3 | 19.55 | 166 | 24.79 | 5208 |

Compared with other methods, Inception V3 model has obvious superiority in evaluating the Top-1 error rate of the discrimination accuracy index [10]. Compared with the existing AlexNet method, the false positive rate of this method is reduced by nearly 1/2. Google introduced the Google Network Model in 2014, which includes five convolutional layers, three pooled layers, one fully connected layer, and 11 initial components [11]. As you can see from Table 1, although the InceptionV3 schema has a network depth of about 20 AlexNet schemas, it has fewer than 38.5 million parameters due to its factored operation [12]. In the selection of teacher model, people should first consider the accuracy rate of pattern recognition, so this paper chooses "cognitive model" as the teacher model [13].

When selecting the student model, people should focus on the number of parameters required and the time required [14]. The number of parameters required is directly related to the storage capacity required, and the time required will also be greatly increased [15]. AlexNet is a deep neural network with 5 convolutional networks and 3 fully connected structures established by Hinton et al., University of Toronto, Canada, in 2012. AlexNet is computationally equivalent to 1/21 of V3 and VGG16. Although there is no special advantage in parameter quantization [16]. This project chooses AlexNet mode as the research object in view of the shortcomings of GPU in computing performance in reality [17].

### B. Data Compression in network Mode

Select InceptionV3, which has a more complex structure and a deeper layer, and AlexNet, which has a simpler structure and a shallower number of layers [18]. First of all, the independent study of the teacher to get a satisfactory effect, and then the study of the teacher [19]. Secondly, students are modeled and softmax is used to convert the obtained learning results into learning methods [20]. so as to enhance the learning effect of the learning object. In Figure 1, the teacher model InceptionV3 is "softened" with temperature parameter T, and then the softened probability distribution - soft target is obtained by softmax transformation (the picture is quoted in Appl.Sci. 2019, 9(10), 1966).

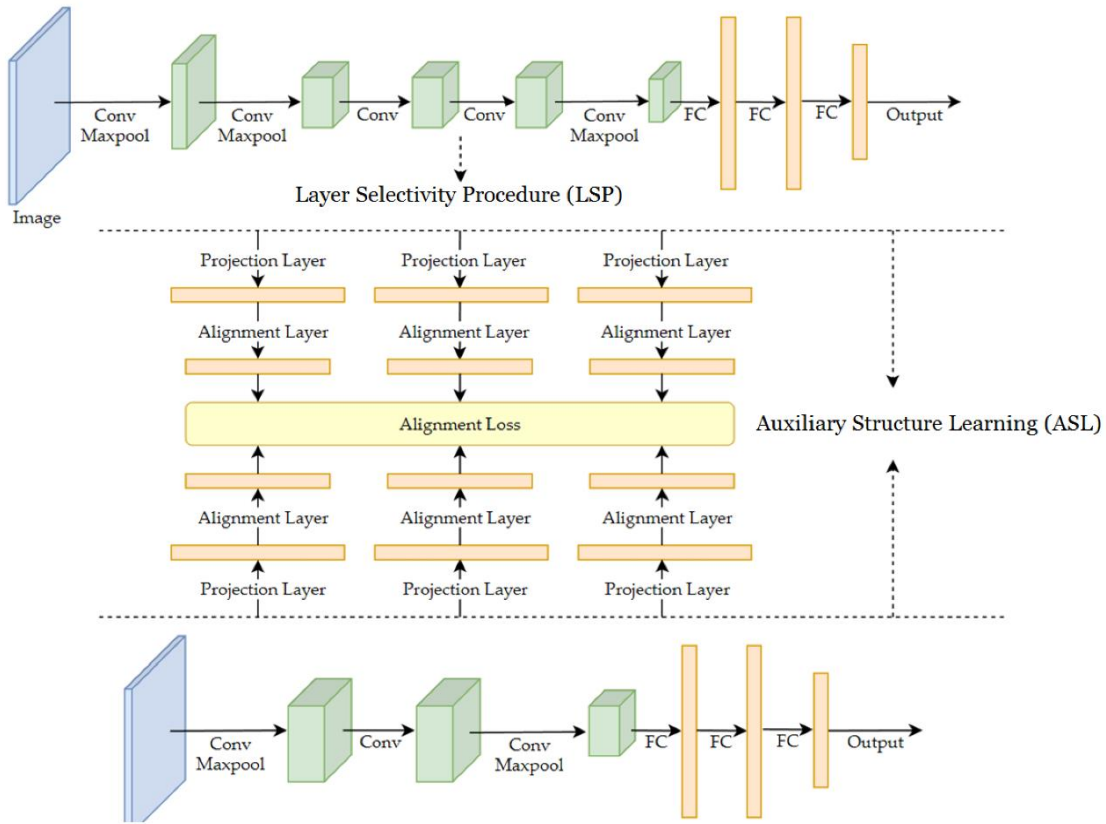

Fig. 1. Knowledge distillation network structure

This project intends to extract the "soft" object from the teacher's modeling, so as to effectively improve its identification performance [21]. In the learning phase, adjusting the value of parameter T is directly related to the learning effect, and usually takes an integer greater than 0. In this study, 1, 10, 20, 30, 40, and 50 were selected as experimental parameters, and their learning effects in T-size (Table 2) were compared [22]. When the parameter is set to 1, the accuracy, specificity, and sensitivity obtained are better than other options, so the adjustment parameter T is set to 1 in future learning [23].

TABLE II. EXPERIMENTAL RESULTS OF DIFFERENT PARAMETERS T

| Parameter T | Accuracy rate /% | loss | Specificity /% | Sensitivity /% |
|---|---|---|---|---|
| 1 | 95.83 | 0.003 | 88.06 | 99.96 |
| 10 | 95.18 | 0.021 | 87.19 | 99.98 |
| 20 | 93.89 | 0.037 | 83.31 | 99.98 |
| 30 | 93.24 | 0.054 | 81.15 | 99.45 |
| 40 | 92.26 | 0.068 | 80.29 | 99.45 |
| 50 | 90.00 | 0.121 | 78.57 | 99.45 |

## C. Classification Process

This paper proposes an idea based on transfer learning, that is, after network training, all the training parameters are saved. The goal is to improve learning outcomes and reduce the probability of overlearning [24]. Initialize and learn parameters at the fully connected level. Finally, the effectiveness of the proposed algorithm is verified by simulation [25]. The method of "teaching" by the teacher is proposed to realize the learning of various parameters required in the modeling process of the teacher [26]. So that the whole small network has strong discernability [27]. Image steganography mainly extracts residual features from images by establishing residual models, and uses statistical models to strengthen features and combines machine learning methods to achieve classification. For an image of length $m$ and width $n$, $U = \{U_{i,j}\} \in R^{m \times n}$, where the pixel value, $U_{i,j}$, ranges from 0 to 255. Embedded information is arbitrary multimedia information, which is converted into bit data stream steganography in an image carrier during embedding. The steganographic information is a matrix of the same size as the image $U$, $E = \{E_{i,j}\} \in R^{m \times n}$, and $E_{i,j} = \{0, -1, +1\}$. The encoding of steganographic information is extracted by comparing the original image and analyzing the location of pixel value change. Spatial adaptive steganography selects steganographic position $Q = \{(i_0, j_0), (i_1, j_1), \cdots, (i_\sigma, j_\sigma)\}$ in an image by designing a distortion function and minimizing the calculated distortion value, where $\sigma$ is the length of a stream of steganographic information bits in an image. The steganographic image is $Z = \{Z_{i,j}\} \in R^{m \times n}, Z = U + E$. For a 3×3 region in the image, the steganographic image consists of

$$\begin{bmatrix} Z_{i-1,j-1} & Z_{i,j-1} & Z_{i+1,j-1} \\ Z_{i-1,j} & Z_{i,j} & Z_{i+1,j} \\ Z_{i-1,j+1} & Z_{i,j+1} & Z_{i+1,j+1} \end{bmatrix} = \begin{bmatrix} U_{i-1,j-1} & U_{i,j-1} & U_{i+1,j-1} \\ U_{i-1,j} & U_{i,j} & U_{i+1,j} \\ U_{i-1,j+1} & U_{i,j+1} & U_{i+1,j+1} \end{bmatrix} + \begin{bmatrix} E_{i-1,j-1} & E_{i,j-1} & E_{i+1,j-1} \\ E_{i-1,j} & E_{i,j} & E_{i+1,j} \\ E_{i-1,j+1} & E_{i,j+1} & E_{i+1,j+1} \end{bmatrix} \quad (1)$$

The algorithm adds two hiding modes of +1 and +1 to the watermark, and its hiding probability is 0, which is the same as the artificially added low frequency noise [28]. The proportion of steganography in the whole image is very small, so it needs to be preprocessed to reduce its interference to the image in order to obtain the densification characteristics. Fridrich et al. established a hidden detection depth modeling - Super resolution (SRM), which extracts the residual information of dense images by establishing the association between edge points and core points to obtain hidden information. The residual $R_{res}$ is calculated by

$$R_{res} = \hat{Z}_{i,j} - \lambda Z_{i,j} \quad (2)$$

$\hat{Z}_{i,j}$ is the pixel, and the $\lambda$ order predicted value of $Z_{i,j}$ is obtained from the pixels around $Z_{i,j}$. The relationships constructed in SRM include simple relationships in the pixel subtractive adjacency model and other complex relationships, respectively

$$R_{res} = Z_{i,j+1} - Z_{i,j}$$
$$R_{res} = (2Z_{i,j-1} - Z_{i-1,j-1} - Z_{i+1,j-1} + 2Z_{i+1,j}) - 4Z_{i,j} \quad (3)$$

Residuals introduce a nonlinear relationship by selecting the smaller or larger value of the two-pixel relationships, thus improving the diversity of the image. For example, the amount of surplus is minimal both horizontally and vertically

$$R_{res} = \min\{(Z_{i,j-1} + Z_{i,j+1} - 2Z_{i,j}), (Z_{i-1,j} + Z_{i+1,j} - 2Z_{i,j})\} \quad (4)$$

In this method, truncation operation is used to define the residual interval, and it is convenient to describe the residual characteristics by co-existing matrix [29]. Quantization operation is used to enhance the hiding effect of watermarking, which makes the residual characteristics of the dense image and the picture have a big difference.

$$R_{res} = trunc_T\left(round\left(\frac{R_{res}}{c}\right)\right) \quad (5)$$

$c$ is the step size of quantization, $c \in \{\lambda, 1.5\lambda, 2\lambda\}$ when $\lambda > 1$, and $c \in \{1, 2\}$; $round(\cdot)$ is an integer operation when $\lambda = 1$. $trunc_T(\cdot)$ is a truncated operation. The residual information extracted was statistically processed by the co-existence matrix. Use classical machine learning algorithms such as integrated classifiers to identify hidden images.

In A pixel region $U' \in R^{t \times t}, t \in \{t \mid t = 2k+1, k \in Z\}$ of the carrier image, if the central pixel is, $U_{i,j}$, then the vector formed by the surrounding pixels is, $M_{i,j}^{U'}$, satisfying the condition, $M_{i,j}^{U'} \subset U', U_{i,j} \notin M_{i,j}^{U'}$. Suppose there is a linear function $f$ which can obtain the C order value of the central pixel according to the surrounding pixels, $\lambda U_{i,j}$, namely

$$f(M_{i,j}^{U'}) = \lambda U_{i,j} \quad (6)$$

After the steganographic information $E'$ is embedded in the carrier image region $U'$, it becomes steganographic image region $Z'$, the center pixel in the steganographic region, and the residual $res \in R$ of $Z_{i,j}$ is obtained by the difference between the surrounding pixels of the center pixel, the center pixels of order $M_{i,j}^{Z'}$ and $\lambda$, and $\lambda Z_{i,j}$ namely

$$R_{res} = R(M_{i,j}^{Z'}, Z_{i,j}) = f(M_{i,j}^{Z'}) - \lambda Z_{i,j} \quad (7)$$

From equation (1), it can be seen that the steganographic image in equation (7), $M_{i,j}^{Z'}, Z_{i,j}$ Be composed of

$$M_{i,j}^{Z'} = M_{i,j}^{U'} + M_{i,j}^{E'}$$
$$Z_{i,j} = U_{i,j} + E_{i,j} \quad (8)$$

Given the linear property of the function $f$

$$R_{res} = f(M_{i,j}^{U'} + M_{i,j}^{E'}) - \lambda(U_{i,j} + E_{i,j})$$
$$= f(M_{i,j}^{E'}) - \lambda E_{i,j} + f(M_{i,j}^{U'}) - \lambda U_{i,j} \quad (9)$$

According to equation (6), the final residual value is

$$R_{res} = f(M_{i,j}^{E'}) - \lambda E_{i,j} \quad (10)$$

After difference processing, the residual property of carrier image is 0, while the residual property of noisy image is steganography. An image segmentation method based on residual error is proposed, which can improve the recognition rate of the object to be detected. Thirty high-pass filters are studied by SRM method [30]. The processing of image convolution is essentially the addition of pixel weights. In this method, first order residuals are extracted from the left edge, and then moved column by column according to the step size. Finally, residuals are obtained by adding each pixel point and corresponding weight, and the residuals are sorted according to the order of their regions to form a residuals map. The procedure for generating the residual Graph is shown in Figure 2 (Graph neural networks: A review of methods and applications).

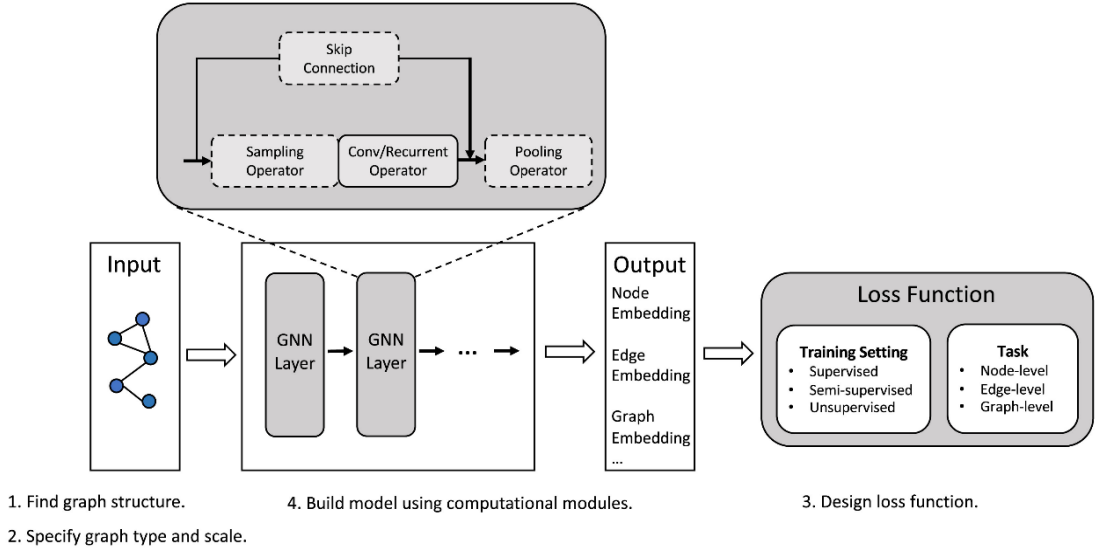

Fig. 2. Generation process of residual graph

A residual filter of size $k \times k$ performs image convolution in an image of size $m \times n$, and when there is no filling and the step size is $s \times s$, the size of the feature graph $R$ is

$$R_{width} = \frac{m-k}{s} + 1$$
$$R_{length} = \frac{n-k}{s} + 1 \quad (11)$$

In the surrounding pixels of the center pixel, the adjacent pixels have the highest correlation with the center pixel, so the adjacent pixels are used to construct the residual extraction function. In order to prevent the introduction of image information and eliminate the excess $Z_{i,j}$ component of $\lambda = \sum_p w_p$, the feature formula is transformed, that is

$$R_{res} = \sum_p w_p (Z_{i,j}^p - Z_{i,j}) \quad (12)$$

Where $Z_{i,j}^p - Z_{i,j}$ is the directional difference. The calculation of the residual can be converted into a linear combination of directional differences, and the parameter $w_p$ of the linear combination is optimized according to the value. Parameter updating is achieved by introducing momentum, and the calculation process in pytorch deep learning framework is

$$w_p' = w_p - v l_k \quad (13)$$

$v$ is the momentum, which is calculated by

$$v = \alpha v' + \frac{\partial L_{loss}}{\partial w_p} \quad (14)$$

III. EXPERIMENTAL DESIGN

A. *Experimental Platform*

This project takes PyTorch as the research object. PyTorch is a software for deep learning and deep neural networks, it uses PyTorch to construct neural networks, without writing all the execution methods, you can easily call the corresponding function library. This will greatly shorten the learning cycle and improve the learning efficiency.

## B. Data collection and preprocessing

The experimental dataset featured medical imagery captured from standard clinical examinations of individuals across various healthcare facilities. These images were made available in 2020 by Dieleman, an investigator affiliated with the University of California, San Diego, on the public repository Mendeley. Comprising 5915 chest X-ray photographs, the collection is stored in JPEG file format. The distribution of data is shown in Table 3.

TABLE III. DISTRIBUTION OF DATA SET COMPONENTS

| Data type | Pneumonia. | Normal | Total |
|---|---|---|---|
| Training set | 3922 | 1363 | 5285 |
| Test set | 394 | 236 | 630 |
| total | 4316 | 1599 | 5915 |

X-ray is the main evidence for clinical diagnosis of pneumonia, and the quality of the image is directly related to the diagnostic effect of the disease. Therefore, the collected image data are first screened by senior doctors to obtain high-quality images, and then the relevant professionals manually mark the images. The X-ray images were divided into two samples, which were manually labeled by two different pulmonary specialists. In order to prevent human labeling errors and ensure the accuracy of identification results, different experts are used for labeling[31][32]. In order to meet the input requirements of the InceptionV3 and AlexNet modes, the size of the X-ray image is adjusted to the sizes of 299x299 and 224x224 pixels (Figure 3). The ratio of the number of X-ray images in the test set to the training set is approximately 1:10. In addition, the denoising problem also needs attention, and Yan's work has provided us with inspiration. By adopting unsupervised learning methods to deal with the denoising problem of Magnetic Resonance Imaging (MRI), Yan's research demonstrated the advanced application of deep learning technology in medical image processing [33]. By using a content encoder and a random noise encoder to separate the content information and noise in the image, and by regularizing the noise distribution through Kullback-Leibler (KL) divergence loss, it provides us with important references for optimizing our knowledge-extraction-based Convolutional Neural Network. Furthermore, combining adversarial loss and cycle consistency loss ensures the consistency of content information between noise input and denoised output images, which directly enlightens us on our goal to improve recognition performance while reducing the consumption of computational resources. Therefore, Yan's methodology not only technically supports our research but also offers us a new perspective on how to effectively utilize deep learning technology for processing medical images .

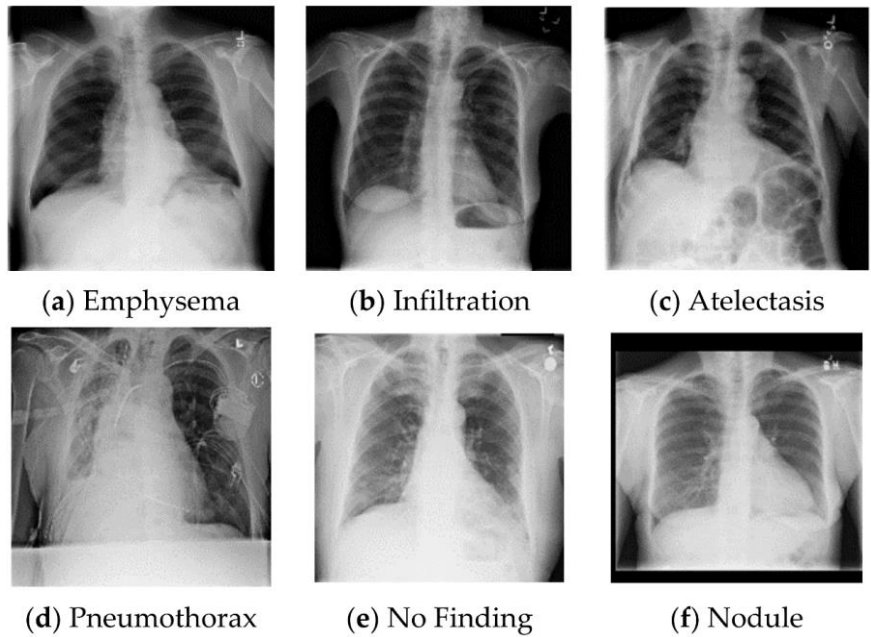

Fig. 3. Sample chest X-ray images

## C. Parameter Settings

During the training of InceptionV3, a method based on transfer learning is used to initialize the parameters of the whole connectivity layer[34][35]. Set the learning rate to $5 \times 10^{-3}$, and set the decay rate to 0.9 and decay once during 7 epochs. With softmax function as the discriminant function and mutual entropy function as the damage function, the SGD results are used to solve the problem[36]. At present, the most common algorithm is Adam algorithm, but this algorithm has a shortcoming, that is, the algorithm convergence speed is too fast, it is difficult to get the best. Through comparative test, the conclusion that SGD method is more effective is drawn. In AlexNet network training, the learning rate is set to $2 \times 10^{-4}$, the attenuation is 0.9, and the transfer learning algorithm similar to V3 is used. The optimization method also adopts stochastic gradient descent algorithm[37].

*D. Experimental process and results*

This paper compares the common deep neural network models. In this model, the loss is a result of the amount of mutual entropy, which reflects the probability distribution between the predicted results and the actual data. Learning loss is often used to guide learning[38]. When it is less than 0.02, it indicates that the classifier has good learning performance. AlexNet, VGG16, ResNet18, ResNet34, Inception V3 were selected in this paper, and their recognition effects on the same training set and test set are shown in Table 4.

TABLE IV. RECOGNITION ACCURACY OF EACH MODEL

| Model name | Accuracy rate /% | loss |
|---|---|---|
| AlexNet | 94.552 | 0.006 |
| VGG16 | 91.646 | 0.021 |
| ResNet18 | 91.802 | 0.005 |
| ResNet34 | 95.469 | 0.015 |
| Inception V3 | 96.333 | 0.024 |

The recognition of the three modes with additional structure is much better in CT images. Therefore, this paper chooses the Teacher module with the initial V3 mode in the optimal algorithm, and takes AlexNet with less network structure as the learning module. The use of these two system resources was compared during the training (Table 5).

TABLE V. SYSTEM RESOURCE USAGE OF ALEXNET AND INCEPTION V3 MODELS

| Model name | Video memory usage /GB | GPU usage /% | Training time /h |
|---|---|---|---|
| AlexNet | 1.89 | 44.44 | 0.10 |
| Inception V3 | 3.54 | 95.96 | 0.81 |

After the training, the difference of the performance of the optimized AlexNet (hereinafter referred to as AlexNet_S) model for pneumonia CT image classification was compared between the AlexNet model before optimization. The results show that the improved algorithm can not only improve the diagnosis of lung diseases, but also improve the specificity and sensitivity of indicators. In contrast, AlexNet_S mode is the student module, which can reduce the memory space by nearly half, which makes it more flexible to configure in a variety of computers with different memory capacities. At the same time, GPU utilization decreased by 51 percent, which allowed Student Module to run on computers with poor GPU performance, greatly improving neural network porting performance (Table 6).

TABLE VI. ALEXNET MODEL PERFORMANCE COMPARISON BEFORE AND AFTER OPTIMIZATION UNIT: %

| Model name | Accuracy rate | specificity | sensitivity | GPU occupancy |
|---|---|---|---|---|
| Teacher Module | 93.41 | 81.59 | 99.78 | 95.96 |
| AlexNet | 91.69 | 80.54 | 99.78 | 44.44 |
| AlexNet_S | 95.83 | 88.06 | 99.78 | 44.44 |

In contrast to the baseline AlexNet architecture, AlexNet_S achieved enhancements in the metric model's classification precision by 4.1 and 2.39 percentage points accordingly. The optimized model particularly excels in the initial detection of pneumonia; however, as its sensitivity rises, there is a concurrent reduction in the misclassification rate for pneumonia cases. Evaluation outcomes indicate that AlexNet_S surpasses the original AlexNet in terms of sensitivity and its sensitivity index closely mirrors that of the reference V3 model. Specificity is a key factor to evaluate the ability of this model to distinguish normal patients, and its high specificity can reduce the diagnostic error rate. In terms of specificity, AlexNet_S and InceptionV3 were 7.45% and 6.41 percentage points, respectively.

IV. CONCLUSION

This paper proposes a method to classify pneumonia CT images using convolutional neural networks. AlexNet and InceptionV3 neural networks with different structure and depth are used to obtain the optimal learning algorithm AlexNet_S which exceed InceptionV3. The enhancements in the AlexNet model, facilitated through the incorporation of knowledge extraction technology, have underscored the potential of deep learning techniques in medical image analysis. This approach not only addresses the limitations associated with deep neural networks' demand for computing resources but also sets a new benchmark for the implementation of efficient and effective diagnostic tools in medical imaging. Furthermore, our findings advocate for the continued exploration and development of optimized CNN models that can operate within the constraints of available computational resources while maintaining high standards of accuracy and efficiency. Such advancements hold the promise of revolutionizing the field of computer-assisted diagnostic imaging, making it more accessible, reliable, and cost-effective. Ultimately, this research paves the way for future studies to further refine these methods and explore their applicability across a broader spectrum of medical imaging tasks, potentially enhancing the diagnostic processes and patient outcomes in the realm of healthcare.